\documentclass[aps,prl,amsmath,amssymb,reprint,nobibnotes,longbibliography,superscriptaddress]{revtex4-2}

\usepackage{t1enc}
\usepackage[utf8]{inputenc}
\usepackage{graphicx}
\usepackage{amsmath}
\usepackage{hyperref}

\begin{document}

\title{500 microkelvin nanoelectronics}

\author{Matthew Sarsby}
\thanks{These authors contributed equally to this work.}
\author{Nikolai Yurttag\"{u}l}
\thanks{These authors contributed equally to this work.}
\author{Attila Geresdi}
\email[Present address: Department of Microtechnology and Nanoscience, Chalmers University of Technology, SE 41296 Gothenburg, Sweden; e-mail: ]{geresdi@chalmers.se}

\affiliation{QuTech and Kavli Institute of Nanoscience, Delft University of
Technology, 2600 GA Delft, The Netherlands.}

\maketitle

\textbf{Fragile quantum effects such as single electron charging in quantum dots
or macroscopic coherent tunneling in superconducting junctions are the basis of
modern quantum technologies. These phenomena can only be observed in devices
where the characteristic spacing between energy levels exceeds the thermal
energy, \boldmath$k_\textrm{B}T$, demanding effective refrigeration techniques
for nanoscale electronic devices. Commercially available dilution refrigerators
have enabled typical electron temperatures in the $10$ to $100\,$mK regime,
however indirect cooling of nanodevices becomes inefficient due to stray
radiofrequency heating and weak thermal coupling of electrons to the device
substrate. Here we report on passing the millikelvin barrier for a
nanoelectronic device. Using a combination of on-chip and off-chip nuclear
refrigeration, we reach an ultimate electron temperature of
$T_\textrm{e}=421\pm35\,\mu$K and a hold time exceeding $85\,$hours below
$700\,\mu$K measured by a self-calibrated Coulomb-blockade thermometer.}

\section*{Introduction}

Accessing the microkelvin regime \cite{pickett2018} holds the potential of
enabling the observation of exotic electronic states, such as topological ordering
\cite{RevModPhys.82.3045}, electron-nuclear ferromagnets
\cite{PhysRevLett.111.147202, chekhovich2013}, p-wave superconductivity
\cite{RevModPhys.75.657} or non-Abelian anyons \cite{stern2010} in the
fractional quantum Hall regime \cite{Nayak_2008}.
In addition, the error rate of various quantum devices, including single
electron charge pumps \cite{RevModPhys.85.1421} and superconducting quantum
circuits \cite{Devoret1169} could improve by more effective thermalization of
the charge carriers.

The conventional means of cooling nanoelectronic devices relies on the heat flow
$\dot{Q}$ between the refrigerator and the conduction electrons, mediated by
phonon-phonon coupling in the insulating substrate, $\dot{Q}_\textrm{p-p}
\propto T_\textrm{p1}^4-T_\textrm{p2}^4$ with $T_\textrm{p1}$ and
$T_\textrm{p2}$ being the respective phonon temperatures, and electron-phonon
coupling  in the device $\dot{Q}_\textrm{e-p} \propto
T_\textrm{e}^5-T_\textrm{p}^5$ where $T_\textrm{e}$ is the electron temperature
\cite{hotelectron,RevModPhys.78.217}, both heat flows rapidly diminishing at low
temperatures. In a typical dilution refrigerator, $T_\textrm{p}$ is larger than $5\,$mK, and
specially built systems reach $1.8\,$mK \cite{cousins1999advanced}, which limits
$T_\textrm{e}$ larger than $T_\textrm{p}$ to be in the millikelvin regime. The lowest static electron temperature
reached with this technique was $T_\textrm{e}=3.9\,$mK \cite{bradley2016nanoelectronic}
in an all-metallic nanostructure, and other experiments reached similar values
\cite{XIA2000491, doi:10.1063/1.3586766, iftikhar2016} in semiconductor
heterostructures.

This technological limitation can be bypassed by adiabatic magnetic
refrigeration, which relies on the constant occupation probability of the energy
levels of a spin system, proportional to $\exp(-g\mu Bm/k_\textrm{B}T)$ in the absence of
heat exchange with the environment \cite{gorter1934}. Here, $k_\textrm{B} T$ is
the thermal energy at a temperature of $T$, $g\mu B$ is the energy split between
adjacent levels at a magnetic field of $B$ and $m$ is the spin index.
Thus, the constant ratio $B/T$ allows for controlling the temperature of the
spin system by changing the magnetic field. Exploiting the spin of the nuclei
\cite{kurti1956nuclear, progress}, this technique has been utilized to cool
nuclear spins in bulk metals down to the temperature range of $T$ around $100\,$pK
\cite{Knuuttila2001}. If only Zeeman splitting is present, which is linear in
$B$, the ultimate temperature is limited by the decreasing molar heat capacity,
$C_\textrm{n}=\alpha B^2/T^2$ in the presence of finite heat leaks,
$\dot{Q}_\textrm{leak}$. Here, the prefactor
$\alpha=N_0I(I+1)\mu_\textrm{n}^2g_\textrm{n}^2/3k_\textrm{B}$ contains $I$, the size of the spin,
$g_\textrm{n}$, the g-factor, $N_0$, the Avogadro number and $\mu_\textrm{n}$, the nuclear
magneton.

\begin{figure*}[!ht]
\centering
\includegraphics[width=\textwidth]{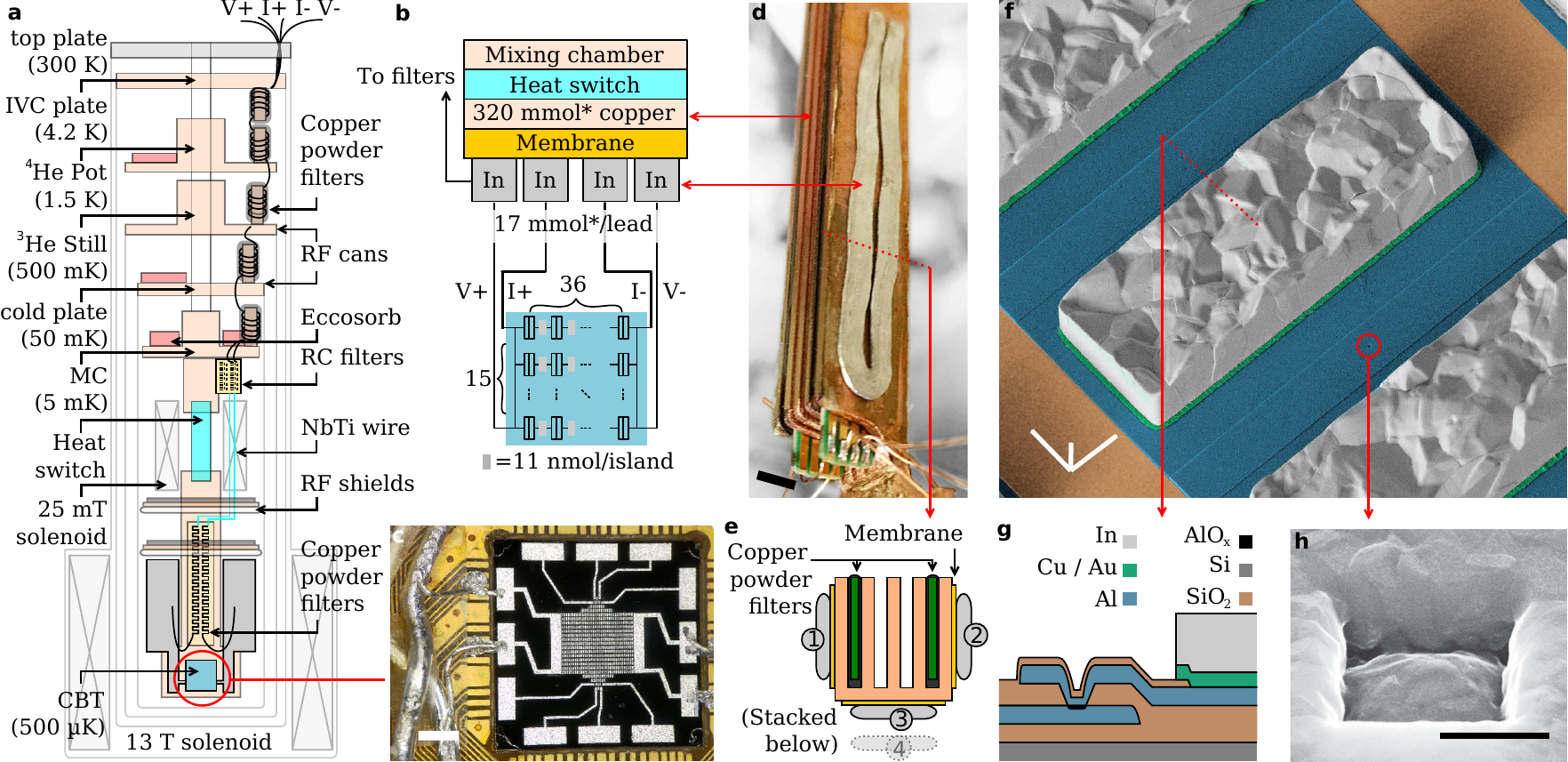} 
\caption{
\textbf{The experimental setup for microkelvin nanoelectronics.}
(a) The cross-sectional sketch of the setup integrated into a dilution
refrigerator with the characteristic temperatures listed on the left. The
device, a Coulomb-blockade thermometer (CBT) with the integrated indium
refrigerant is shown as a cyan box, encircled in red. (b) The thermal path
between the mixing chamber of the dilution refrigerator and the CBT (highlighted
in cyan, the grey blocks denote In cooling fins). The electrical signal lines
pass through the indium blocks. The effective molar amount of the refrigerant
material is listed in each block, respectively. (c) Photograph of the CBT chip
with the press-welded connections to the off-chip indium stages consisting of
$1\,$mm diameter indium wires are visible on the left. (d) Photograph of the
indium refrigerant mounted on the copper block with an electrically insulating
membrane in between. The picture was taken during the assembly of the stage. The
slots of the copper block contain the copper powder filters of the electrical
lines to reduce radiofrequency heating, see also the cross-sectional image in
panel (e). (f) False-coloured scanning electron micrograph of an island of the
CBT, with a cross-sectional schematics shown in panel (g) along with the colour
legend. The encircled Al/AlO$_x$/Al tunnel junction is shown in higher
magnification in panel (h). The scale bars denote $2\,$mm (c), $5\,$mm (d),
$20\,\mu$m (f) and $500\,$nm (h), respectively. See the Data Availability
section for raw images.}
\end{figure*}

Bulk nuclear cooling stages have predominantly been built of copper
\cite{progress}, owing to its high thermal conductivity, beneficial metallurgic
properties and weakly coupled nuclear spins, which allows for magnetic
refrigeration of the nuclear spins down to $50\,$nK \cite{PhysRevLett.42.1702},
however the weak hyperfine interaction results in a decoupling of the electron
system at much higher temperatures, around $1\,\mu$K in bulk samples. Another
material, a Van Vleck paramagnet, PrNi$_5$, has been used as a bulk nuclear
refrigerant exploiting its interaction-enhanced heat capacity in a temperature
range above $200\,\mu$K \cite{Folle1981} even in dry dilution refrigerators
\cite{1367-2630-15-11-113034}.

Integrating the nuclear refrigerant with the nanoelectronic device yields a
direct heat transfer between the electrons and nuclei $\dot{Q}_\textrm{e-n}=\alpha
\kappa^{-1} B^2\left(T_\textrm{e}/T_\textrm{n}-1\right)$ per mole. Here, $\kappa$ is the Korringa
constant, and $T_\textrm{e}$, $T_\textrm{n}$ are the electron and nuclear spin temperatures,
respectively. Owing to an $\alpha/\kappa$ ratio $60$ times better than that of
copper, indium has recently been demonstrated as a viable on-chip nuclear
refrigerant \cite{yurttagul2018indium}. In addition, indium allows for on-chip
integration of patterned thick films by electrodeposition and has been
demonstrated as a versatile interconnect material for superconducting quantum
circuits \cite{rosenberg2017}.

However, on-chip nuclear refrigeration is limited by the large molar heat leaks
$\dot{Q}_\textrm{leak}$, which thus far limited the attainable electron
temperatures above $3\,$mK both for copper
($\dot{Q}_\textrm{leak}=0.91\,$pW mol$^{-1}$, \cite{bradley2017chip}) and indium
($\dot{Q}_\textrm{leak}=0.69\,$pW mol$^{-1}$,\cite{yurttagul2018indium}) as on-chip
refrigerant, with a total heat leak of approximately $4\,$pW in both of these
experiments. Furthermore, only very short hold times were attained with the
chips warming up during the magnetic field sweep, limiting the practical
applications of these devices. In contrast, encapsulating the chip in a
microkelvin environment has led to superior cooling performance using copper,
yielding $T_\textrm{e}=2.8\,$mK with a hold time of approximately $1\,$hour below $3\,$mK
\cite{palma2017and}.

Here, we report on a combined on- and off-chip nuclear demagnetization approach,
which surpasses the $1\,$mK barrier in a
nanoelectronic device. We demonstrate an electron temperature below $500\,\mu$K
by combining on-chip indium cooling fins with an off-chip parallel network of
bulk indium leads attached to a copper frame. By demagnetization cooling of all
components together we greatly reduce the heat leak to the nanoscale device,
enabling both a record low final electron temperature and long hold times in
excess of $85\,$hours, demonstrating the viable utilization for quantum
transport experiments in the microkelvin regime.

\section*{Results}

\subsection*{Setup for on- and off-chip indium nuclear demagnetization}

We discuss the experimental setup following the cross-sectional drawing in
Fig.~1a. Our design is accommodated by a commercial dilution refrigerator
(MNK126-700, Leiden Cryogenics) with several microwave-tight radiation shields to reduce
electronic noise coupling to the nanostructure. The magnetic field required for
magnetic cooling was applied by a superconducting solenoid with a rated field of
$13\,$T. The device was installed in the field centre, however parts of the
off-chip refrigerants were subjected to smaller magnetic fields defined by the
field profile of the solenoid, which we account for by listing the effective
molar amount of the materials.

We attach the copper frame ($1.7\,$mol total, $0.32\,$mol effective amount) to
the mixing chamber via an aluminium-foil heat switch
\cite{doi:10.1063/1.1135452}, which is activated by a small solenoid at $\approx
10\,$mT of applied magnetic field. Crucially for our design, the electrical measurement lines are not
connected to the copper stage, rather to the parallel network of nuclear cooled
indium wires with a $23\,$mmol total and $17\,$mmol effective amount per line.
These wires are attached to the copper frame by an epoxy resin membrane
(nominal thickness of $40\mu$m, Fig.~1b) which provides an additional layer of thermal isolation during nuclear
cooling. The electrical and thermal contact to the device is made by press
welding the indium wires onto electroplated indium bonding pads on the chip
(Fig.~1c).

We apply both electronic filtering and microwave shielding in order to reach
sub-1 mK electron temperatures. We thermalize all electrical lines by
custom-made copper-powder filters \cite{thalmann2017comparison} at each stage of
the dilution refrigerator and a fifth order RC filter with a cutoff frequency of
$50\,$kHz at the mixing chamber. To decouple the cold chip below $1\,$mK from
the thermal noise of the mixing chamber \cite{glattli1997noise}, we install an
additional set of copper-powder filters within the copper stage
which is demagnetizated together with the indium lines (Fig.~1d,e). Further
details on the measurement circuit are listed in Supplementary Figures 1-4, and in the Methods.

\begin{figure}[!ht]
\centering
\includegraphics[width=0.5\textwidth]{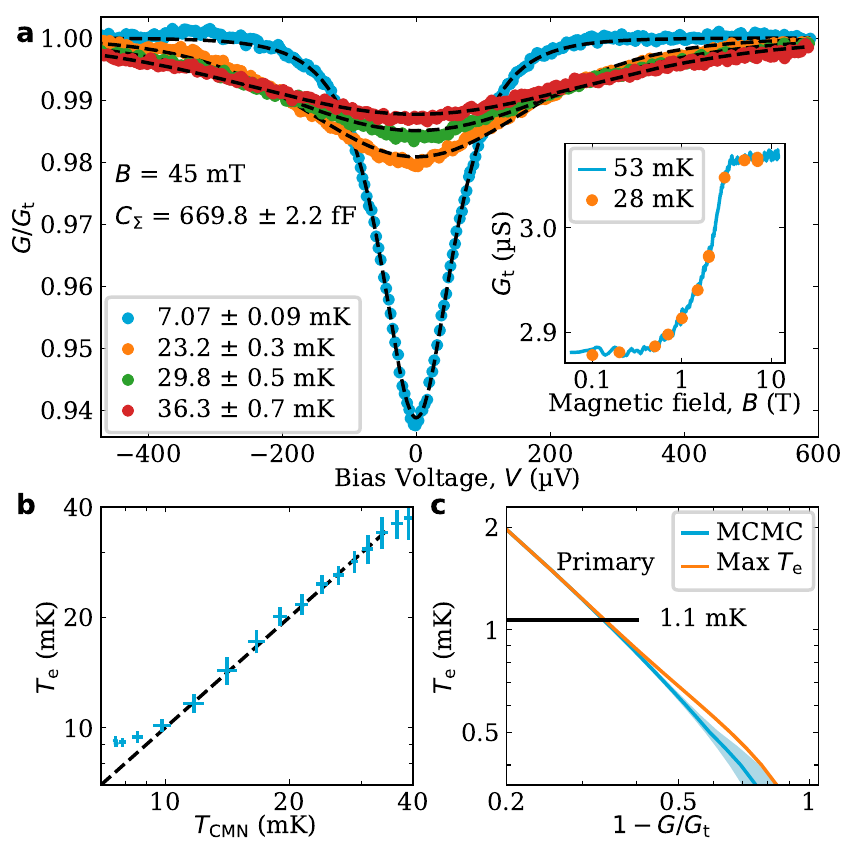}
\caption{\textbf{Calibration and primary operation of the CBT.}
(a) The voltage-dependent differential conductance, $G$ of the device at
different temperatures. The conductance is normalized with the conductance at
large bias, $G_\textrm{t}$ and fitted against the single electron tunneling
model to determine the electron temperature for each curve and the total
capacitance per island, $C_\Sigma=669.8\pm2.2\,$fF. The inset shows
$G_\textrm{t}$ as a function of the magnetic field, at two distinct
temperatures. (b) The electron temperature, $T_\textrm{e}$ at $B=100\,$mT with
the heat switch closed as a function of the mixing chamber temperature of the
dilution refrigerator, measured by a calibrated cerium magnesium nitrate (CMN)
mutual inductance thermometer. The temperature errors are shown as horizontal
and vertical lines, respectively. (c) The theoretical zero bias calibration
curve based on $E_\textrm{C}$ with the
$k_\textrm{B}T_\textrm{e}=0.4E_\textrm{C}$ limit \cite{feshchenko2013primary}
depicted as the horizontal line, see text. The theoretical maximum of
$T_\textrm{e}(G/G_\textrm{t})$ with all CBT islands in full Coulomb blockade is
denoted by the orange line. The average $T_\textrm{e}$ and the $3\sigma$
confidence interval are shown as the blue solid line and the blue shaded region,
respectively, based on the Markov chain Monte Carlo (MCMC) method, see section
Integrated Coulomb blockade thermometry. See the Data Availability section for
raw data.}
\end{figure}

\subsection*{Integrated Coulomb blockade thermometry}

To directly measure the electron temperature, we utilize a Coulomb-blockade
thermometer (CBT) \cite{Pekola1}. CBTs rely on the universal suppression of
charge fluctuations in small metallic islands enclosed between tunnel barriers
at low electron temperatures. With a total capacitance of each island,
$C_\Sigma$, we define the charging energy for a tunnel junction array of length
$N$ to be $E_\textrm{C}=e^2/C_\Sigma \times (N-1)/N$ with $e$ being the charge of a
single electron. In the weak charging regime, $k_\textrm{B}T_\textrm{e}\gg E_\textrm{C}$, the full
width at half maximum of the charging curve is $eV_{1/2} = 5.439 N k_\textrm{B}
T$, providing the possibility of primary thermometry independent of $E_\textrm{C}$
defined by the device geometry \cite{Farhangfar1}. Crucially for
temperature measurements during demagnetization cooling, CBTs exhibit no
sensitivity to the applied magnetic field \cite{doi:10.1063/1.367397}.

We fabricated a $36\times 15$ array with an ex-situ Al/AlO$_x$/Al tunnel
junction process \cite{prunnilaexsitu, yurttagul2018indium} to create junctions
of $780\times780\,$nm$^2$ in size (Fig.~1h).
Since the highly resistive tunnel junctions with $R_\textrm{j}=35\,\textrm{k}\Omega$
correspond to a high thermal barrier, we electrodeposited indium cooling fins
($50\times 140\times 25.4\,\mu$m$^3$ corresponding to $11.4\,$nmol, see
Fig.~1f,g) on each island for local nuclear cooling. We note however that the measured
device has only $5$ conducting lines, resulting in an $N\times M=36\times 5$ CBT
array. The details of the device fabrication are previously described in
\cite{yurttagul2018indium}.

In Fig.~2a, we show the normalised differential conductance $G(V)/G_\textrm{t}$ of the
CBT at different temperatures obtained by conventional low frequency lock-in
technique at $19.3\,$Hz. We use the master equation of single electron tunneling
\cite{Pekola1,Farhangfar1} to find the electron temperature of each curve and
the charging energy as a global fit parameter, $E_\textrm{C}=232.6\pm0.8\,$neV$=k_\textrm{B}\cdot2.7\,$mK, yielding
$C_\Sigma=670\pm2\,$fF. Remarkably, at a small magnetic field of $B=45\,$mT
required to suppress superconductivity, we achieve an electron temperature
$T_\textrm{e}=7.07\pm0.09\,$mK by phonon cooling, attesting to the well-shielded
environment of the device. At this low temperature, we had to account for the
Joule heating of the CBT chip \cite{kautz1993self}, see the Supplementary
Note 1 and Supplementary Figure 6. Upon calibration, we use the zero bias conductance $G(V=0)$ for
continuous thermometry, see the Supplementary Note 1. We note that we
observe a temperature-independent $\Delta G_\textrm{t}/G_\textrm{t}\approx 0.03$ magnetoresistance
between zero field and $12\,$T (inset of Fig.~2a), which we account for during
the magnetic field ramps. To demonstrate the limitations of phonon cooling, we
evaluate $T_\textrm{e}$ as a function of the mixing chamber temperature and observe a
saturation behavior below $10\,$mK (Fig.~2b).

In the low temperature regime, where $E_\textrm{C}$ is larger than $k_\textrm{B}T_\textrm{e}$, random island
charge offsets influence the conductance of the CBT
\cite{feshchenko2013primary}, which causes an a statistical uncertainity of the
inferred $T_\textrm{e}$. To estimate this confidence interval, we use the following
numerical procedure. First, we calculate $G/G_\textrm{t}$ by a Markov chain electron
counting model with a random set of offset charges for each island. Using the
Markov chain Monte Carlo (MCMC) approach, we evaluate the conductances of the
tunnel junction chain for 1000 gate charge configurations at a given $T_\textrm{e}$.
Finally, we build a histogram of the total CBT conductance, which is the a sum
of $M=5$ randomly selected line conductances. The conclusion of our calculations
is that the conductance distribution remains surprisingly narrow for our array
consisting of 35 islands in series and 5 rows in parallel and enables a
temperature measurement with a $3\sigma$ relative confidence interval of less
than $10\%$ for $T_\textrm{e}$ above $400\,\mu$K$=0.15E_\textrm{C}$. We plot the calculated calibration
curve $T_\textrm{e}(G/G_\textrm{t})$ (blue solid line) together with a $3\sigma$ confidence
interval in Fig.~2c (blue shade), which provides the quoted statistical error of
$T_\textrm{e}$. We also display the $T_\textrm{e}=0.4E_\textrm{C}/k_\textrm{B}=1.1\,$mK limit of the
universal behaviour \cite{feshchenko2013primary}, where offset charges do not
play a role in the device conductance. In addition, we plot (orange curve) the
theoretical maximum of $T_\textrm{e}(G/G_\textrm{t})$, which occurs when all islands are in full
Coulomb blockade. For the MCMC calculation, we assumed uniform tunnel junction
resistances and island capacitances. Our fabrication process yields a junction
resistance variation of $\delta R/R$ less than $10\%$ \cite{yurttagul2018indium}, which
yields an upper bound of systematic temperature error of $\Delta T_\textrm{e}/T_\textrm{e}=k(\delta
R/R)^2$ around $1\%$, where $k$ is a geometry-dependent prefactor of the order of
unity \cite{hirvi1996numerical}. The details of our numerical simulations can be
found in the Supplementary Note 1.

\begin{figure}[!ht]
\centering
\includegraphics[width=0.5\textwidth]{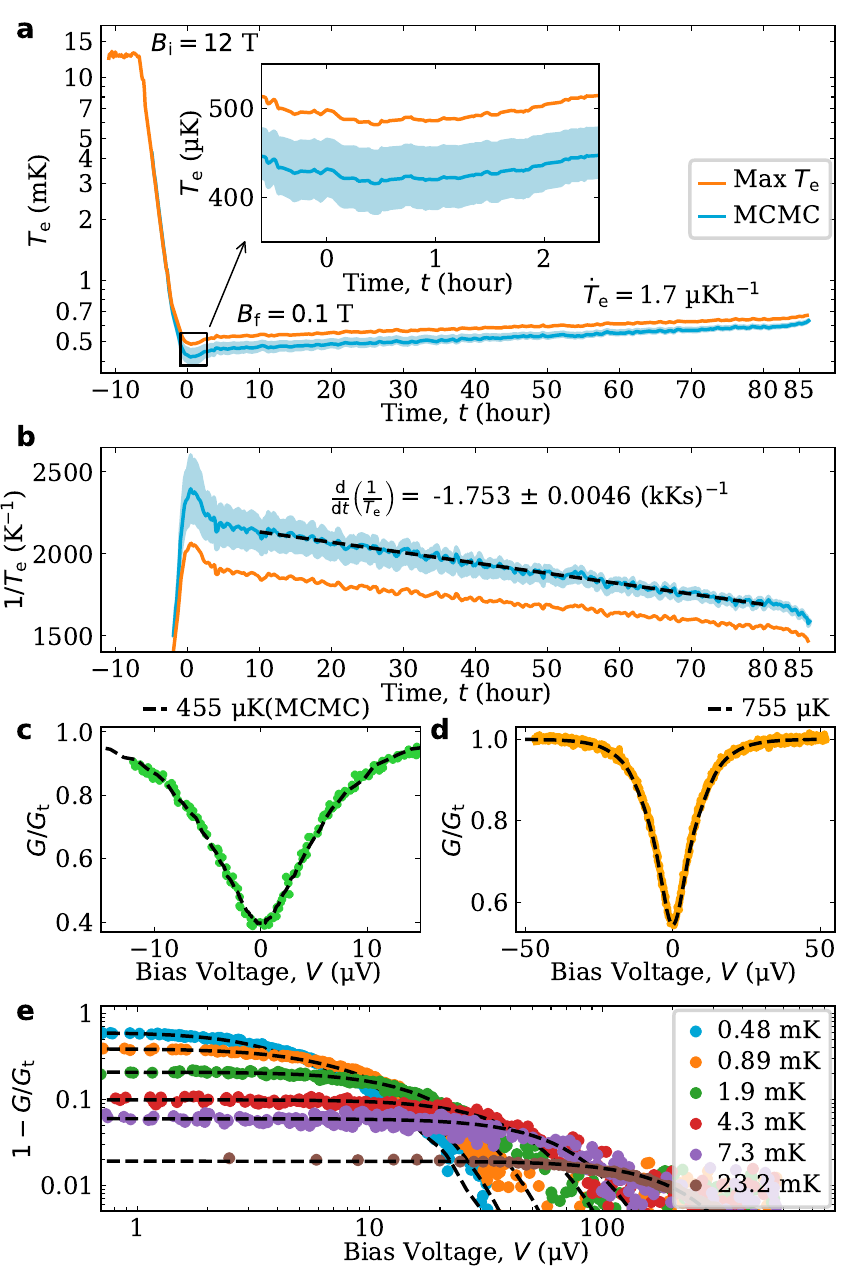}
\caption{\textbf{Nuclear demagnetization of the CBT to the microkelvin regime.}
(a) A nuclear demagnetization experiment starting from $T_\textrm{e}=13\,$mK and
$B_\textrm{i}=12\,$T ramping down to $B_\textrm{f}=0.1\,$T with a ramp rate of
$\dot{B}(B)=[-2\times 10^{-4}\textrm{s}^{-1}] \cdot B$.
The inset shows a closer view of the minimum electron temperature reaching well
below $T_\textrm{e}=500\,\mu$K for all calibrations, see text. (b) The inverse
temperature $1/T_\textrm{e}$ as a function of time, which is used to estimate the heat
leak, see text. Panels (c) and (d) display finite bias voltage data in
demonstration of the primary CBT operation in the microkelvin regime. (e) Finite
voltage bias data of the CBT in a broad temperature range. In panels (c), (d)
and (e), the experimental data is shown as circles. The dashed lines are the
fitted theoretical curves based on the master-equation model for $T_\textrm{e}$ above $1\,$mK
and on the charge-offset averaged MCMC model for lower temperatures, see the section Integrated Coulomb blockade thermometry and the Supplementary Note 1.
The inferred $T_\textrm{e}$ values are listed in the legends. See the Data Availability section for raw data.}
\end{figure}

\subsection*{Electron cooling by nuclear demagnetization}

We initialize a nuclear demagnetization experiment by precooling the device with
a closed heat switch at $B_\textrm{i}=12\,$T and reach $T_\textrm{e,i}=13\,$mK after precooling
for $164\,$hours. After opening the heat switch, we remove the magnetic field
with a decreasing rate of $\dot{B}(B)=[-2\times 10^{-4}\textrm{s}^{-1}] \cdot B$
to $B_\textrm{f}=0.1\,$T, which results in a decreasing $T_\textrm{e}$ (Fig.~3a). After finishing
the ramp, we find the lowest measured electron temperature $T_\textrm{e}=421\pm35\,\mu$K
calibrated using our MCMC model, averaged over a period of one hour (see inset
of Fig.~3a). The quoted $3\sigma$ confidence interval is based on the
conductance variation calculated by the MCMC model (see Fig.~2c). We note that
the full blockade limit of the CBT also yields $T_\textrm{e,min}$ less than $500\,\mu$K (orange
line).

After reaching the minimum temperature, the device gradually warms up. We find a
warm-up rate of $\dot{T}_\textrm{e}=1.7\,\mu$K h$^{-1}$, and a hold time of $85\,$hours with
$T_\textrm{e}$ less than $700\,\mu$K. This duration was limited by the periodic cryogenic liquid
transfer to the host cryostat, which induced mechanical vibrations and
consequently a rapid warmup of the device.

As a proof of concept quantum transport experiment in the
microkelvin regime, we measure the finite bias charging curve of the
CBT at two different temperatures below $1\,$mK (circles in Fig.~3c,d).
Remarkably, we find an excellent agreement with the calculated curves, using
$T_\textrm{e}$ as the single fit parameter (dashed lines). In addition, we observe no
distortion of the lineshape due to Joule heating at finite voltage biases,
further demonstrating the role of strong electron-nucleus coupling in indium. To
demonstrate the wide range of primary thermometry in our regime of interest, we
display a set of experimental data taken between $T_\textrm{e}=480\,\mu$K and $23.2\,$mK
(Fig.~3e) together with the fitted theoretical curves on a logarithmic
scale (circles and dashed lines, respectively).

\begin{figure}[!ht]
\centering
\includegraphics[width=0.5\textwidth]{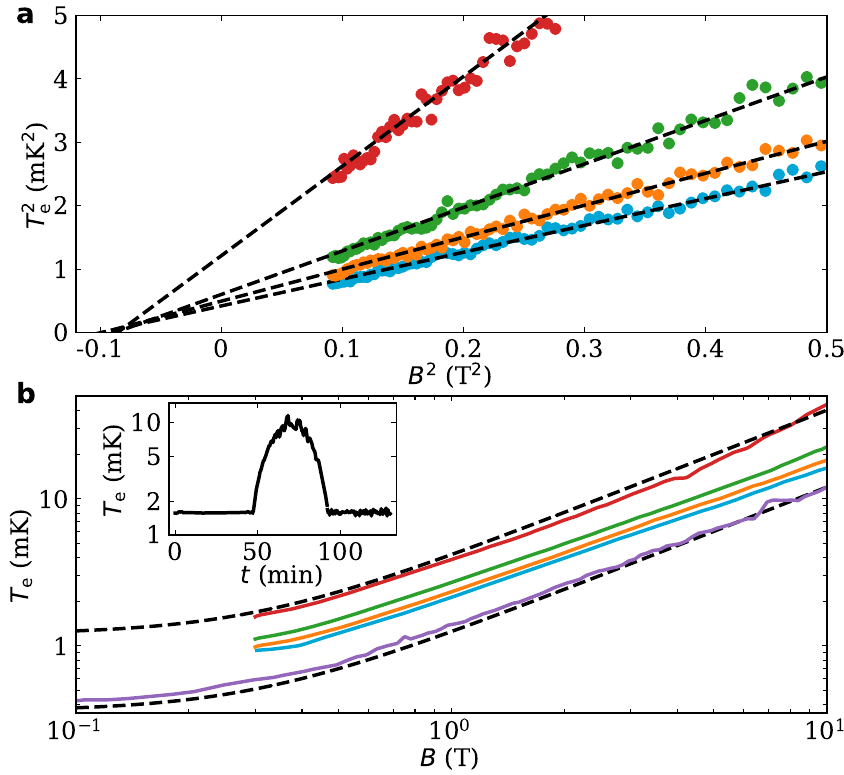}
\caption{\textbf{Adiabaticity of indium as a nuclear refrigerant for electrons.}
(a) The intercept of the extrapolated $T_\textrm{e}^2(B^2)$ linear regression
(dashed lines) with different starting conditions define the internal magnetic field of
$|b|=295\pm7\,$mT. The experimental data is shown as circles.
(b) The $T_\textrm{e}(B)$ plots demonstrate that the internal field limits the
obtained electron temperature. Experimental data is shown as solid lines. The
dashed lines with constant $T/\sqrt{B^2+b^2}$ assuming perfect adiabaticity are
shown with two different starting conditions. The inset shows a magnetic field
ramp $B=1\,\textrm{T}\rightarrow8\,\textrm{T}\rightarrow1\,\textrm{T}$ yielding an
adiabatic efficiency of $\eta=1.00\pm0.04$, see the section Internal magnetic field and heat leak calculation. See the Data Availability section for raw data.}
\end{figure}

\subsection*{Internal magnetic field and heat leak calculation}

Next, we turn to the refrigeration properties of indium, and display a series
of $T_\textrm{e}(B)$ curves in Fig.~4. Notably, $T_\textrm{e}^2(B^2)$ follows a linear
relation with a negative intercept, $-b^2=-(295\pm7\,$mT$)^2$ (Fig.~4a),
which demonstrates that indium does not follow a constant $B/T$ ratio. We
attribute our results to the strong quadrupolar splitting due to the
inhomogenuous electric field in the crystalline lattice \cite{pobellmatter}. A
resulting effective magnetic field $b$ then adds to the applied magnetic field,
$B_\textrm{tot}^2=B^2+b^2$ \cite{pobellmatter}, which limits the lowest
attainable temperature, even if $B\ll b$ (Fig.~4b). Our measurements yield
similar $b$ values to earlier nuclear demagnetization experiments in bulk indium
samples \cite{Tang1}. This correspondence directly confirms that the measured
$T_\textrm{e}$ and the inferred nuclear spin temperature, $T_\textrm{n}$ are close to each other,
which is the key requirement for nuclear demagnetization cooling of
nanoelectronics.

Based on the obtained internal magnetic field, we estimate the heat leak of the
CBT using the warm-up rate $\textrm{d}T_\textrm{n}/\textrm{d}t=\dot{Q}_\textrm{leak}/C_\textrm{n}$, where $C_\textrm{n}=n
\alpha B_\textrm{tot}^2/T_\textrm{n}^2$ is the nuclear spin heat capacity with $n$ being
the molar amount of the refrigerant. These equations yield
$\dot{Q}_\textrm{leak}= - n \alpha
B_\textrm{tot}^2\frac{\textrm{d}(1/T_\textrm{e})}{\textrm{d}t}$, where
$\textrm{d}(1/T_\textrm{e})/\textrm{d}t$ is the inverse warmup rate (Fig.~3b) in the
limit of $T_\textrm{n}=T_\textrm{e}$. At $B=100\,$mT, we find
$\dot{Q}_\textrm{leak}=26.7\,$aW per island, which is lower than the $\dot{Q}_\textrm{leak}=32\,$aW per island obtained
earlier using copper \cite{palma2017and}.

The small heat leak is further attested by the direct demonstration of the
adiabaticity of the experiment. In the inset of Fig.~4b, we show a measurement
run starting with $B_\textrm{i}=1\,$T and $T_\textrm{e,i}=1.57\pm0.01\,$mK. After ramping up the
magnetic field up to $8\,$T and back to $B_\textrm{f}=1\,$T, we acquire
$T_\textrm{e,f}=1.57\pm0.06\,$mK. With $B_\textrm{f}=B_\textrm{i}$, the adiabatic efficiency
\cite{palma2017and} reads $\eta=T_\textrm{e,i}/T_\textrm{e,f}=1.00\pm0.04$, unity within error
margin. This remarkable result demonstrates the outstanding control of $T_\textrm{e}$ by
setting the magnetic field, which is unprecedented for nuclear magnetic cooling
stages built for nanoelectronics.

\section*{Discussion}

In conclusion, we demonstrated nuclear magnetic cooling of electrons in a
nanostructure down to an ultimate temperature of $T_\textrm{e}=421\pm35\,\mu$K.
Integrating on- and off-chip refrigeration utilizing the strong electron-nucleus
coupling of indium, we achieve $85\,$hours of hold time in the microkelvin
range, which opens up this, so far unexplored, temperature regime for
nanoelectronic devices and quantum transport experiments in pursuit of 
topological and interacting electron systems. We note that other transition
metals with a large $\alpha/\kappa$ ratio \cite{pobellmatter} can potentially be
used in future experiments to target an ultimate temperature regime of
$T_\textrm{e}\approx T_\textrm{n}\sim \mu_\textrm{n} g_\textrm{n} b/k_\textrm{B}$ determined by the internal
magnetic field $b$. However, comparing our results with earlier experiments
featuring on-chip-only refrigeration \cite{bradley2017chip, yurttagul2018indium}
demonstrates the importance of a cold electronic environment provided by
off-chip refrigeration \cite{palma2017and}. Furthermore, due to the finite heat
leaks, on-chip refrigeration is required to effectively cool highly resistive
devices, where the thermalization via the leads is inefficient. In addition, we
extend the range of primary nanoelectronic thermometry below $1\,$mK, opening
avenues for metrological applications and quantum sensing in the ultra-low
electron temperature regime. Cooling down electrons to the microkelvin regime
also paves the way towards investigating the limits of quantum coherence in
solid state devices on the macroscopic scale \cite{RevModPhys.90.025004}.

\section*{Methods}

\subsection*{Mechanical design}

The copper stage is milled from 5N5 purity copper. It is left unannealed to
reduce eddy currents. The connection to the heat switch is a single M4 bolt, with both the
mating surfaces gold plated. The single heat switch in this design resides in a field
compensated region of the main magnet. It is formed with a welded lamination of
aluminium foil, connecting to copper foil at each end.
The aluminium foil layers are mechanically deformed removing continuous paths
for superconducting vortices. The switch is then encapsulated in epoxy laced with
copper powder for stiffness and filtering. A small superconducting solenoid
surrounds the heat switch.

The heat switch shows closing behaviour at an applied
current of $\pm90\,$mA through the solenoid, without measurable
dependence on the main field.
We observe an abrupt change in both the mixing chamber and CBT temperatures when
the solenoid is energised.
To be confident of a closed switch, even at high fields, we choose a working
current of $140\,$mA to close the switch.
The superconducting lines to the the switch quenched when operated above approximately $300\,$mA.
The heat switch current was sourced from a TTi EL301R Power Supply (PSU) running in
constant current mode.
The copper magnet leads are filtered and thermalised with winding embedded in
copper powder epoxy at the $4\,$kelvin plate before transitioning to
superconducting NbTi. The room temperature wiring is wrapped around a ferrite
ring for additional filtering. However, when the heat switch power supply is
attached, the galvanic isolation to the cryostat from the measurement
electronics is broken, so for best noise performance the leads to the PSU are
physically unplugged when performing the nuclear cooling experiments.

The molar amount of material in the copper stage is calculated to be $1.7\,$moles.
However, as not all of the material is in the field centre, the contribution to the heat capacity and
nuclear cooling reduces accordingly. 
We calculate the effective molar amount using the field profile of the main
solenoid along the vertical axes, $B(y)$ and the cross-sectional areas, $A(y)$.
The effective molar amount is then 
\begin{equation}
n^\star = \frac{1}{V_\textrm{m}} \int \left(\frac{B(y)}{B_\textrm{max}}\right)^2
A(y) \mathrm{d}y = 0.32\,\textrm{mol},
\end{equation}
where $V_\textrm{m}$ is the molar volume and $B_\textrm{max}$ is the field maximum.

Similarly, we estimate the effective amount of the indium lines to be
$17\,$mmol with a total amount of $23\,$mmol per line.

The indium stages are made from 6N $2\,$mm indium wire rolled out
with a small non-ferrous metal rolling pin to flatten it onto a thin
epoxy membrane on the copper stage. From there, 4N $1\,$mm indium
wires are stir welded at room temperature to ensure good low temperature interconnects.
Connecting to the chip, we used 4N $0.7\,$mm indium wires, which
are welded to the $1\,$mm wires and press fit onto the indium
plated bond pads on the chip. We anneal the indium in-situ over 2 days at room
temperature using a turbo molecular pump before cooling the experiment down.

The electrical lines for measuring the CBT connect from the mixing chamber to
the microkelvin stage through NbTi superconducting lines with the sorrounding
copper matrix removed using dilute nitric acid.

\subsection*{Zero bias tracking}

During the magnetic field sweeps and warm-up, we infer the electron temperature
based on the zero bias conductance of the CBT. However, time-dependent bias
voltage drifts can offset the device, resulting in a systematic error in the
measured electron temperature. We note that this error source increases as the
temperature is lowered due to the increasing curvature of the conductance dip.
To mitigate this effect, we continuously tracked the device conductance in a
small voltage bias window centered around the conductance minimum, which was
monitored on-the-fly. Note the visible drift in Supplementary Figure 5. The quoted
temperatures are then extracted from the conductance minima of a rolling
second-order polinomial fit. This procedure removes the systematic positive
temperature error introduced at finite voltage biases, but allows for
the temperature tracking during demagnetization.

\subsection*{Filtering of the electronic lines}

We paid attention to condition the RF environment around the chip,
installing RF absorbing materials, in the volumes between plates to
reduce the resonant quality factor of the chambers, and covered inner surfaces
with an infrared absorbing paint. A set of copper grills are added onto the
nuclear cooled stage to damp microwave resonances without increasing eddy
current heating in the microkelvin stage.

We used copper powder filters at all stages of the dilution refrigerator. The
mixture is 1:3 by mass of copper to epoxy, which also matches the thermal
expansion of copper. Voids in the epoxy casting were removed in a vacuum
chamber. As the epoxy, we used  Loctite Stycast 2850FT with catalyst CAT 24LV,
chosen for lower viscosity. The copper powder was supplied by ALDRICH Chemistry, and
specified as copper powder(spheroidal), $20\pm10\,\mu$m, 99\% purity. We note
that we used the same mixture in all stages of the setup.

We used a combined RC and copper powder filter anchored to the mixing chamber
plate of the dilution refrigerator. The RC filter component provides a sharp
cutoff at $50\,$kHz. The entire board is covered by copper powder epoxy to
prevent microwave leakage. The board accommodates 12 lines in arranged in 6
pairs.

The RC filter components are listed below:\\
SMD multilayer ceramic capacitors, 50603 [1608 Metric] $\pm5\%$ C0G/NP0.\\
Parts: C0603C152J5GACTU, C0603C332J5GACTU, C0603C472J5GACTU\\
SMD Chip Resistor, 0402 [1005 Metric], $1.5\,$k$\Omega$, ERA2A Series, 50 V,
Metal film.\\ Parts: ERA2AEB751X, ERA2AED152X.

In addition, we further thermalize the experiment with home-made 
filters mounted inside the slots of the copper nuclear cooled stage.
These are made from printed circuit boards with a long meandering trace
resulting in an effective continuous element filter. We measure
$R=4.6\,\Omega$ line resistance and a $C\approx 250\,$pF
capacitance to electrical ground defined by the copper stage. The PCBs have four
meandering copper traces on each side.
The dual copper layers and the $0.8\,$mm board thickness slot into
$1\,$mm spaces in the copper stage. The gaps are then filled with
copper powder epoxy.

\section*{Corresponding author}
Correspondence and request of materials should be addressed to Attila Geresdi.

\section*{Acknowledgements}
The authors thank J.~Mensingh, O.~Benningshof, N.~Alberts,
R.~N.~Schouten and C.~R.~Lawson for technical assistance.
This work has been supported by the Netherlands Organization for
Scientific Research (NWO) and Microsoft Corporation Station Q.
A.~G.~acknowledges funding from the European Research Council under the European
Union's Horizon 2020 research and innovation programme, grant number 804988.

\section*{Author contributions}
A.~G.~designed and supervised the experiments.
N.~Y.~designed and fabricated the CBT devices with integrated indium fins.
M.~S.~designed and fabricated the off-chip nuclear cooling stages and filters.
M.~S.~ and N.~Y.~performed the experiments. N.~Y.~implemented the numerical CBT
array simulation. All authors analyzed the data. The manuscript has been
prepared with contributions from all the authors.

\section*{Competing interests}
The authors declare no competing interests.

\section*{Data availability}
Raw datasets measured and analyzed for this publication are available at the
4TU.ResearchData repository, DOI:
\href{https://dx.doi.org/10.4121/uuid:ffaeb9fc-9baf-428e-8a33-7e4b451d8f9e}{10.4121/uuid:ffaeb9fc-9baf-428e-8a33-7e4b451d8f9e} (Ref.~\cite{rawdata}).

\def\bibsection{}

\section*{References}

\bibliography{references}
\bibliographystyle{naturemag}

\newpage

\end{document}